\RequirePackage{rotating}
\documentclass[a4paper,fleqn,usenatbib]{mnras}

\usepackage{aas_macros}
\usepackage{epsfig}
\usepackage{verbatim, amsmath, amsfonts, amssymb,amssymb}
\usepackage{listings}
\usepackage[utf8x]{inputenc}
\usepackage[table]{xcolor}
\usepackage{adjustbox}
\usepackage{afterpage}
\usepackage{textcomp}
\usepackage{lscape}
\usepackage{natbib}
\usepackage{url}

\newcommand{\iout}[2]{
\begin{itemize}
\item \textbf{input:} #1
\item \textbf{output:} #2
\end{itemize}
}

\title[SN\,Ia spectroscopic diversity with DRACULA]{Exploring the spectroscopic diversity of type Ia supernovae with DRACULA: a machine learning approach}

\author[Sasdelli et al.]
{
M. Sasdelli$^{1,2}$\thanks{E-mail: m.sasdelli@ljmu.ac.uk (MS)},
E. E. O. Ishida$^{2,3}$, 
R. Vilalta$^{4}$,
M. Aguena$^{5}$, 
V. C. Busti$^{5}$,
\newauthor
H. Camacho$^{5}$,
A. M. M. Trindade$^{6,7}$, 
F. Gieseke$^{8}$, 
R. S. de Souza$^{9}$,
\newauthor
Y. T. Fantaye$^{10}$, 
and P. A. Mazzali$^{1,2}$,
for the COIN collaboration
\\
$^{1}$Astrophysics Research Institute, Liverpool John Moores University, Liverpool L3 5RF, UK\\
$^{2}$Max-Planck-Institut f{\"u}r Astrophysik, Karl-Schwarzschild-Stra\ss e 1, 85748, Garching, Germany\\
$^{3}$Clermont Universit\'e, Universit\'e Blaise Pascal, CNRS/IN2P3, Laboratoire de Physique Corpusculaire, BP 10448, F-63000 \\ Clermont-Ferrand, France\\
$^{4}$Department of Computer Science, University of Houston, 4800 Calhoun Rd., Houston TX 77204-3010, USA \\
$^{5}$Departamento de Física Matemática, Instituto de Física, Universidade de São Paulo, CP 66318, CEP 05508-090, \\ São Paulo - SP, Brazil\\
$^{6}$Instituto de Astrofisica e Ciências do Espaço, Universidade do Porto, CAUP, Rua das Estrelas, PT4150-762 Porto, Portugal\\
$^{7}$Departamento de Física e Astronomia, Faculdade de Ciências, Universidade do Porto, Rua do Campo Alegre 687, PT4169-007 \\ Porto, Portugal\\
$^{8}$Institute for Computing and Information Sciences, Radboud University Nijmegen, Toernooiveld 212, 6525 EC \\ Nijmegen, Netherlands\\
$^{9}$MTA E\"otv\"os University, EIRSA ``Lendulet'' Astrophysics Research Group, Budapest 1117, Hungary\\
$^{10}$ Department of Mathematics, University of Rome Tor Vergata, Rome, Italy
}

\date{Accepted XXX. Received YYY; in original form ZZZ}

\pubyear{2016}

\begin{document}
\label{firstpage}
\pagerange{\pageref{firstpage}--\pageref{lastpage}}
\maketitle





\begin{abstract}
The existence of multiple subclasses of type Ia supernovae (SNeIa) has been the subject of great debate in the last decade. One major challenge inevitably met when trying to infer the existence of one or more subclasses is the time consuming, and subjective, process of subclass definition.  In this work, we show how machine learning tools facilitate identification of subtypes of SNe\,Ia through the establishment of a hierarchical group structure in the continuous space of spectral diversity formed by these objects. Using Deep Learning, we were  capable of performing such identification in a 4 dimensional feature space (+1 for time evolution), while the standard Principal Component Analysis barely achieves similar results using 15 principal components. This is evidence that the progenitor system and the explosion mechanism can be described by a small number of initial physical parameters.
As a proof of concept, we show that our results are in close agreement with 
a previously suggested classification scheme and that our proposed method can grasp the main spectral features behind the definition of such subtypes. 
This allows the confirmation of the velocity of lines as a first order effect in the determination of SN\,Ia subtypes, followed by 91bg-like events. Given the expected data deluge in the forthcoming years, our proposed approach is essential to allow a quick and statistically coherent identification of SNeIa subtypes (and outliers).
All tools used in this work were made publicly available in the Python package \texttt{DRACULA} (Dimensionality Reduction And Clustering for Unsupervised Learning in Astronomy)
and can be found within COINtoolbox (\url{https://github.com/COINtoolbox/DRACULA}).
\end{abstract}
\maketitle

\begin{keywords}
    supernovae: general -- methods: machine learning, data analysis, statistical
\end{keywords}


\section{Introduction}
\label{sec:intro}

Type Ia supernovae (SNe\,Ia) are extremely bright objects, exhibiting
a good degree of spectroscopic and photometric homogeneity. 
Among other characteristics, the fact that their luminosity correlates with a set of distance independent quantities constructed using multi-band light curves is particularly relevant. These correlations enable the use of SNe\,Ia as standard candles, which  combined with their strong luminosity, allows us to probe large cosmological distances. This played a major contribution to the discovery of the accelerating expansion of the Universe in the late 20$^{\rm th}$ century  \citep{1998AJ....116.1009R,1999ApJ...517..565P}.

Although most SNe\,Ia are spectroscopically quite uniform,  there is a significant fraction of spectroscopically peculiar objects \citep{2001ApJ...546..734L}, some of which are very different from the average SN\,Ia. At this moment, it is still unclear if there exists different subclasses in the space of SN\,Ia spectra that are truly distinct \citep[e.g][]{2005ApJ...623.1011B} or if subtypes defined in the literature are just extremes of a continuum distribution of properties \citep[e.g][]{2012AJ....143..126B}. 

Parallel to such considerations derived from the analysis of observed spectra, theoretical developments also investigate multiple hypotheses to explain SN\,Ia diversity. A large number of possible progenitor systems and explosion mechanisms have been proposed \citep[e.g.][]{2013FrPhy...8..116H} leading to an agreement that the origin of the majority of SNe\,Ia lies in the thermonuclear runaway of a CO white dwarf in a binary system. Nevertheless, the nature of the companion and the explosion mechanism are still fiercely debated. Proving the existence of well-defined and distinct subclasses would strongly support the hypothesis of different progenitor systems or, qualitatively, different explosion mechanisms.

To identify which spectral feature(s) might carry the signature of physically distinct subclasses (if they exist), a number of different  classification schemes have been proposed. SNe\,Ia have been classified into High and Low Velocity Gradient based on the time gradient of the velocity of the Si~{\sc ii}\,6355\AA\ line \citep{2005ApJ...623.1011B}.
They have also been classified into Shallow and Broad-Silicon classes according to the Equivalent Width (EW) of the Si~{\sc ii}\,6355\AA\ line, referred to as Cool when the ratio between the  Si  {\sc ii}\, 5972\AA\ and the  Si~{\sc ii}\,6355\AA\ is above a certain value at $B$-band maximum \citep{2009PASP..121..238B}.
Many of these classification schemes do not exhibit a clear transition between its subsets when applied to a large number of observations \citep{2012AJ....143..126B}, suggesting that SNe\,Ia characteristics are more a continuum of features than a discretely separable parameter space. However, most of these schemes are based on a very small subset of spectral features. The situation is likely to be quite different if all the information contained in the spectra is taken into account. Additionally, SNe\,Ia are classified in spectroscopic subclasses after the first peculiar object of a certain kind \citep[e.g.][]{2001ApJ...546..734L}. For example, 91T-like when they are similar to SN\,1991T before maximum, 91bg-like when they are similar to the faint SN\,1991bg, and 02cx-like when they are similar to the faint and hot SN\,2002cx. This is a non-quantitative criterion that complicates the study of subtype definition (see appendix \ref{ap:glossary} for a detailed description of the jargon used throughout this paper, including the difference between classes and types).

One main driving force behind the development of classification schemes based on individual spectral features was  the difficulty in obtaining a large number of high quality observations. Having only a few observed objects of  each category, the only viable approach was to minutely study the observations at hand,
extrapolating their characteristics to the entire SN\,Ia population. 
Nowadays, the situation is rapidly changing. In the last few years, data released from a number of observation campaigns increased the number of
available spectra by at least an order of magnitude \citep{2012AJ....143..126B,2012MNRAS.425.1789S, 2013ApJ...773...53F}. In this new paradigm, we face a different challenge: to develop methods and tools capable of dealing with a large number of spectra at once. The overwhelming volume of data defies dependence on human inspection of individual spectra; the process must be automatized. 

Fortunately, similar problems are at the core of machine learning research; such tools can be adapted to a large variety of tasks, as has been reported in other fields \citep[see e.g.][]{crisci2012,libbrecht2015,vidyasagar2015}. Following this trend, the present work is an additional effort 
to popularize modern machine learning techniques within astronomy \citep[see][and references therein]{ball2010,krone-martins2014,ivezic2014}. 

In what follows, we describe a series of machine learning tools and demonstrate how they can help automatize the visualization and classification of a large set of SN\,Ia spectra.  Our goal is to provide a proof of concept, showing that the algorithm is able to leverage the same set of spectral features one would choose by visual recognition, opening the path for an automatic first screening in a situation where the number of available spectra far outnumbers the capacity of the researcher to individually analyse them.
Our approach involves two steps:  reducing the dimensionality of an initially very large space, and subsequently using unsupervised learning (clustering) to automatically identify subtypes of SNe Ia. 
On each step we use state of the art machine learning techniques, which lead to powerful insights on questions underlying SN\,Ia spectral features. The tools used here are implemented in the \texttt{DRACULA} Python package (Dimensionality Reduction And Clustering for Unsupervised Learning in Astronomy) and are publicly available within the COINtoolbox\footnote{\url{https://github.com/COINtoolbox}}.

This paper is organized as follows: Section~\ref{sec:data} describes the data used for our analysis; Section~\ref{sec:DR} explains our approach to dimensionality reduction;
Section \ref{sec:TL} demonstrates how transfer learning can be used in the context of SN\,Ia spectral analysis; Section \ref{sec:comp_ps} shows the improvement in dimensionality reduction achieved by Deep Learning in comparison with Principal Component Analysis; Section \ref{sec:vis} gives a brief overview of the methods used for data visualization;
Section \ref{sec:cluster} reviews the K-Means algorithm; 
Section~\ref{sec:results} goes over our main results. Lastly, Section \ref{sec:discussion} gives summary and discussion. In order to avoid confusion between similar expressions with distinct meanings in the machine learning and astronomy communities, we provide a small glossary in Appendix \ref{ap:glossary} with the definitions used throughout the paper. Appendix \ref{sec:code} describes the \texttt{DRACULA} package, where our proposed tools are implemented.

\section{Data}
\label{sec:data}

We compiled a set of publicly available SN\,Ia spectra from a variety of sources: the Berkeley Supernova Program \citep{2012MNRAS.425.1789S}, the CfA
spectroscopic release \citep{2012AJ....143..126B} and the Carnegie Supernova
Project (CSP) \citep{2013ApJ...773...53F}. Spectra have been collected from the
SUSPECT\footnote{\url{http://www.nhn.ou.edu/~suspect}} \citep{barbon1990, mazzali1995, patat1996, turatto1996, gomez1998, turatto1998, jha1999, li1999, capellaro2001, li2001, salvo2001, hamuy2002, branch2003, valentini2003, benetti2004, garavini2004, anupama2005, gerardy2005, kotak2005, chornock2006, eliasrosa2006, altavilla2007, garavini2007, gerardy2007, hicken2007, krisciunas2007, leonard2007, pastorello2007, phillips2007, stanishev2007, matheson2008, pignata2008, taubenberger2008, wang2008, bufano2009, yamanaka2009, wang2009b} and the WISEREP
\citep{2012PASP..124..668Y} repositories.
CSP spectra are published in rest frame; the remaining spectra were deredshifted using heliocentric redshifts from \cite{2012AJ....143..126B}.

In order to build the input data matrix, the spectra need to be smoothed, binned in a homogeneous wavelength window,  and systematics must be taken into account. Here we follow the procedure used by  \cite{2015MNRAS.447.1247S}, smoothing the spectra through the use of the Savitzky-Golay filter  \citep{morrey1968determining} and applying the derivative over wavelength to the logarithm of the measured flux.  The Savitzky-Golay filter  is effective in reducing a large amount of the noise and, at the same time,  preserving the shape of the features present in the spectra. The use of derivative spectroscopy allows us to remove the systematics due to the uncertainty in the distance determination and in the global spectrum calibration. 
\cite{2015MNRAS.447.1247S} show that the intrinsic luminosity information is well included in the derivative and that there is no significant loss of information. This is confirmed by the study of \cite{2016MNRAS.tmp..696S}. 
The correlation between the luminosity and the spectral features is largely due to the effect of temperature in the behaviour of the spectral lines \citep[e.g.][]{1995ApJ...455L.147N,2006MNRAS.370..299H}.

A possible alternative to the Savitzky-Golay filter and  derivative approach is the use of a wavelet decomposition \citep[see e.g.][]{madgwick2003, paykari2014}, discarding the coefficients heavily affected by reddening and noise \citep{2011MNRAS.414.1617A}. We plan to investigate this further in future work.

Once the pre-processing is done, it is necessary to design the data matrix which will be given as input to the dimensionality reduction algorithm, taking into account the drastic changes in SN spectra with time and the non-ideal epoch coverage. Time sampling of SN data 
is highly irregular, having large periods without observations, specially at very early and very late epochs. In \cite{2015MNRAS.447.1247S} this problem was dealt with by concatenating spectra along the same line in the data matrix, thus taking into account the time evolution of each object. This strategy presents promising results, but generates a matrix with a large fraction of missing data. We propose an alternative approach that allows us to exploit all available spectroscopic information (regardless of the epoch of observation) to attain a stable low dimensional space (Section \ref{sec:TL}). However, before addressing the effectiveness of our proposal, we review main concepts behind dimensionality reduction techniques.

\section{Dimensionality Reduction}
\label{sec:DR}

After the data have been pre-processed, the first step is to transform it to a low dimensional feature space, that is a space of parameters describing well the original input space. We briefly describe below the two main dimensionality reduction 
algorithms used in this work: Principal Component Analysis (PCA) and Deep Learning (DL). 

\subsection{Principal Component Analysis}
\label{subsubsec:PCA}

PCA is a method designed to reduce the dimensionality of a multivariate dataset, by projecting the data onto a lower dimensional feature space. Given its versatility, PCA and variations of it  have been applied to a broad range of astronomical studies \citep[e.g.,][]{yip2004a, yip2004b, ferreras2006, ishida2011a, mitra2011, ishida2011b,graur2013, ishida2013,herrera2013, desouza2014,deSouza2014b, 2015MNRAS.447.1247S}.

The principal components (PCs) are computed diagonalizing the covariance matrix ($\Sigma^2$),  with the eigenvectors being the PCs and the eigenvalues indicating the fraction of total variance {\it explained} by their corresponding  PCs.
The first eigenvector (PC1 - the component associated with the largest eigenvalue) indicates the direction of greatest variance, the second eigenvector (PC2 - component with second largest eigenvalue)  points to the second  direction holding highest variance subjected to being orthogonal to PC1, and so on.

Mathematically, this can be described as follows: given $\Gamma$ 
measured features
$y_1, \ldots, y_{\Gamma}$, all of them column vectors of dimension $n$ (1 for each object in the data set), the first PC is obtained by finding a unit vector $\mathbf{a}$ that maximizes the variance, $S$, of the data projected onto it:
\begin{equation}
\mathbf{a_1} =\underset{||\mathbf{a}||=1}{\rm \arg\max}~ S^2 (\mathbf{a}^ty_1,\cdots,\mathbf{a}^ty_{\Gamma}), 
\label{eq:PC1}
\end{equation}
where $t$ is the transpose operation and   $\mathbf{a_1}$ is the direction of the first PC and $\underset{y}{\arg\max}~  f(y)$ is the set of values of $y$ for which the function $f(y)$ attains its largest value. Once we have computed the $(k-1)^{\rm th}$ PC, the direction of the $k^{\rm th}$ component, for $1 < k \leqslant \Gamma$, is given by 
\begin{equation}
\mathbf{\mathbf{a}_k} = \underset{||\mathbf{a}||=1,\mathbf{a}\bot \mathbf{a}_1,\cdots,\mathbf{a}\bot \mathbf{a}_{k-1}}{\rm \arg\max}S^2(\mathbf{a}^ty_1,\cdots,\mathbf{a}^ty_{\Gamma}), 
\end{equation}
where the condition that each PC must be orthogonal to all previous PCs ensures a new uncorrelated basis. 

It is possible to show that the above is equivalent to computing the eigenvalues and eigenvectors of the $\Sigma^2$ \citep[][chapter 1]{jolliffe1986}. Once the PCs are computed, one can use the percentage of total variance encoded in the eigenvalues in order to determine the dimensionality of the new feature space \citep[see Section 2 of][]{ishida2011a}. 

\subsection{Deep Learning}
\label{subsec:dl}

\begin{figure}
\vspace*{2mm}
\hspace*{1mm}
\includegraphics[width=1.0\columnwidth]{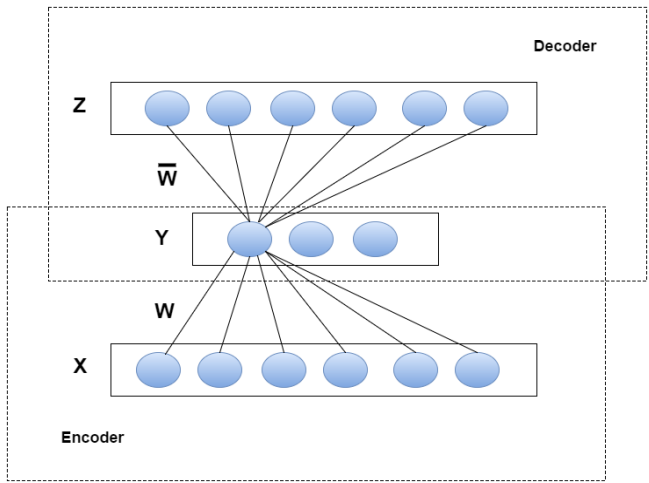}
\vspace*{1mm}
\caption{A simple auto-encoder where the input $\mathbf{X}$ is reproduced in the output layer $\mathbf{Z}$. The middle layer $\mathbf{Y}$ ``compresses'' the input signal $\mathbf{X}$,
effectively reducing the dimensionality of the data.}
\label{autoencoder}
\end{figure}

In Deep Learning (DL), we take the input data $\mathbf{x} = x_1, x_2, ..., x_n$ and represent it in the form of a layer of nodes, or neurons, where each node is a variable $x_i$ (see Fig.~\ref{autoencoder}, bottom layer). Additional layers of neurons above the original input signal are built to ensure that each new layer captures a more abstract representation of the original input signal. In DL, each layer constructs new features by forming non-linear combinations of the features in the layer below. This hierarchical approach to feature construction has been effective in disentangling factors of variation in the data \citep{Hinton06,Bengio13,LeCun15}. DL has contributed to a rapid advancement in the field of neural networks through new mechanisms to train architectures made of many layers of intermediate neurons. 

To illustrate these ideas, consider the task of image processing. Specifically, assume we have an image with thousands of pixels where we wish to recognize the object depicted in the image. We can represent the entire image as a feature vector $\mathbf{p} = (p_1, p_2, ..., p_n)$, where each pixel $p_i$ is a measured  feature. The resulting space is not only very large; in addition, each pixel contains low-level information about the main object in need of recognition. In DL we make each pixel $p_i$ stand as one node along the first layer of the neural network. The second layer is made of nodes computing non-linear combinations of all nodes (pixels) below. The third layer captures non-linear combinations of the nodes on the second layer, and so on. Each layer captures more abstract, global structures of the object under analysis. Starting with low-level pixel information, upper layers can gradually capture edges, motifs, and larger structures of the main object.

While different approaches exist to deal with DL architectures, we focus our attention on the problem of dimensionality reduction. In such unsupervised learning setting, auto-encoders have played a prominent role \citep{Vincent08}. An illustration of a simple auto-encoder is shown in Fig.  \ref{autoencoder}. The goal here is to compress the input signal by a transformation that reduces the size of the feature space. Specifically, the first layer corresponds to input vector $\mathbf{x} \in \mathcal{R}^n$. The intermediate layer corresponds to a new vector $\mathbf{y} \in \mathcal{R}^{d}$, $d < n$, where each node computes a non-linear combination of the input features. The last layer maps the internal representation back to the original dimensionality through a new vector $\mathbf{z} \in \mathcal{R}^n$, with the objective of reproducing the input vector $\mathbf{x}$ as best as possible, $\mathbf{x} \sim \mathbf{z}$. The weight parameters of the auto-encoder are the weight matrices $\mathbf{W}$ and $\mathbf{\overline{W}}$ connecting nodes from one layer to the one above. Weights are optimized during the training phase using the training sample.  We assume weight matrices contain both weights and bias terms (see Fig.~\ref{autoencoder}).  The network is trained by adjusting the weights to minimize an error function that computes the distance between input and output: $\parallel\mathbf{x} - \mathbf{z}\parallel^2$. Specifically, the transformation from the first layer to the second layer ``encodes'' the input signal through a non-linear transformation:

\begin{equation}
\mathbf{y_i} = f_i(\mathbf{x}) = \sigma(\mathbf{w_i}^t \mathbf{x}) 
\end{equation}

\noindent
where $\mathbf{y_i}$ is one intermediate node, $\mathbf{w_i}$ is the weight vector (containing
the bias term), and $\sigma(u)$ can vary in 
nature, a common choice being the sigmoid function $\sigma(u) = \frac{1}{1 + e^{-u}}$.  
The new vector $\mathbf{y}$ is then ``decoded'' into a new vector $\mathbf{z}$ that reconstructs the input vector $\mathbf{x}$:

\begin{equation}
\mathbf{z_j} = g_j(\mathbf{y}) = \sigma(\mathbf{\overline{w_j}}^t \mathbf{y}) 
\end{equation}

While learning to reproduce an input signal may appear as a trivial exercise, the interesting part of the auto-encoder is that the intermediate layer (vector $\mathbf{y}$) ``abstracts'' the representation of the input layer during the training phase, essentially compressing the input data through a combination of non-linear representations. This is similar to the goal behind PCA, except here the combination of features is non-linear, and there is no orthogonality constraint. Each
of the intermediate nodes stands as a new variable in the reduced dimensionality space. 

Notice that the auto-encoder can be divided into two sections. The ``encoder'' section generates more compact representations (Fig.~\ref{autoencoder}; layers 1 and 2), while the ``decoder'' section simply unfolds the compact representation in an attempt to reproduce the
input signal (Fig.~\ref{autoencoder}; layers 2 and 3).

\begin{figure}
\hspace*{2mm}
\vspace*{2mm}
\includegraphics[width=1.0\columnwidth]{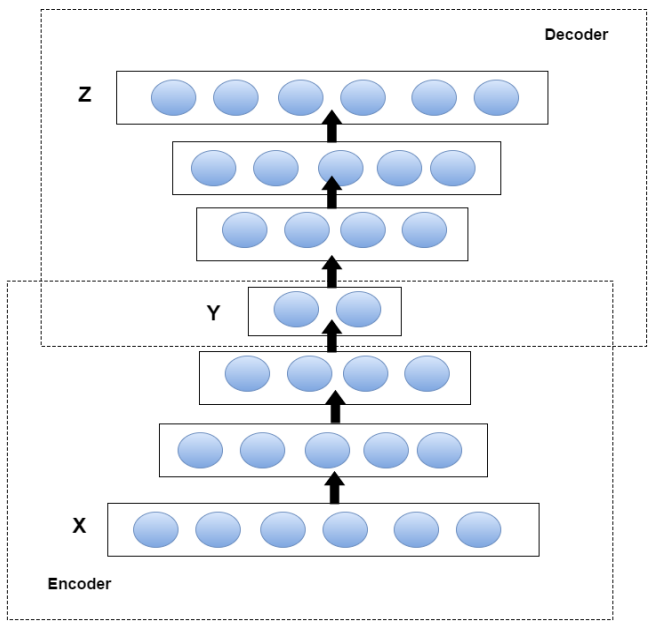}
\vspace*{1mm}
\caption{A deep auto-encoder where intermediate layers provide increasingly more 
abstract representations of the input signal.
The most abstract representation is stored in layer 4 (vector $\mathbf{Y}$).}
\label{deepautoencoder}
\end{figure}

\subsubsection{Stacking Multiple Layers}

The ideas above have been extended to ``deep'' architectures with many layers of neurons. Fig. \ref{deepautoencoder} shows an example of a deep auto-encoder. Training such deep architectures can be done iteratively by stacking several auto-encoders in such a way that the intermediate layer of nodes becomes input to the next auto-encoder. Alternatively we can simply build a deep auto-encoder directly, and let the optimization phase (gradient descent) look for the weight values that minimize the distance between input $\mathbf{x}$ and output $\mathbf{z}$. The net result is a vector $\mathbf{y}$
(middle layer) that in effect summarizes the input signal in a compact fashion. This technique was recently used by \citet{company2015} in the construction of a galaxy morphology catalogue, but its potential in different areas of astronomy still needs to be discovered. This work represents the first effort to use DL techniques in the characterization of spectral features. 

It is important to emphasize that unlike the eigenvectors from PCA, the elements  of the middle layer within a deep network do not follow a natural ordering. 
The power of DL resides on providing a compact representation of the initial data, preserving relevant information through multiple 
layers of abstraction. This compact representation has extraordinary potential in characterizing the data, as will be made clear shortly.

\section{Transfer learning}
\label{sec:TL}

We are interested in a data driven approach to investigate the diversity of SNe\,Ia at maximum brightness. Our strategy consists in reducing the dimensionality of the initial feature space and subsequently applying an unsupervised learning algorithm. However, if we follow the traditional approach of constructing the input matrix only with spectra at maximum (or in a certain epoch bin around it), we will end up with a very small matrix ($\sim$150 objects). It would be difficult for any dimensionality reduction algorithm to grasp the details of a complex space starting with such a small matrix. Concatenating spectra according to the observed epoch bin is a good alternative, but requires a dimensionality reduction tool armed to cope with missing data, as has already been demonstrated in  \citet{2015MNRAS.447.1247S}.

Here we choose to use \emph{transfer learning}, a recent area in machine learning that deals with the general problem of exploiting information from a variety of different environments to help with learning, inference, and prediction in
a new environment where training data is scarce \citep{quionero2009}. 
A simple example is spam filter detection, where one could aim at using the feedback (i.e., labeled data) of a group of existing users to help generating a model for a new user \citep{pan2010}.

\begin{figure}
\includegraphics[scale=0.3, width=1\columnwidth]{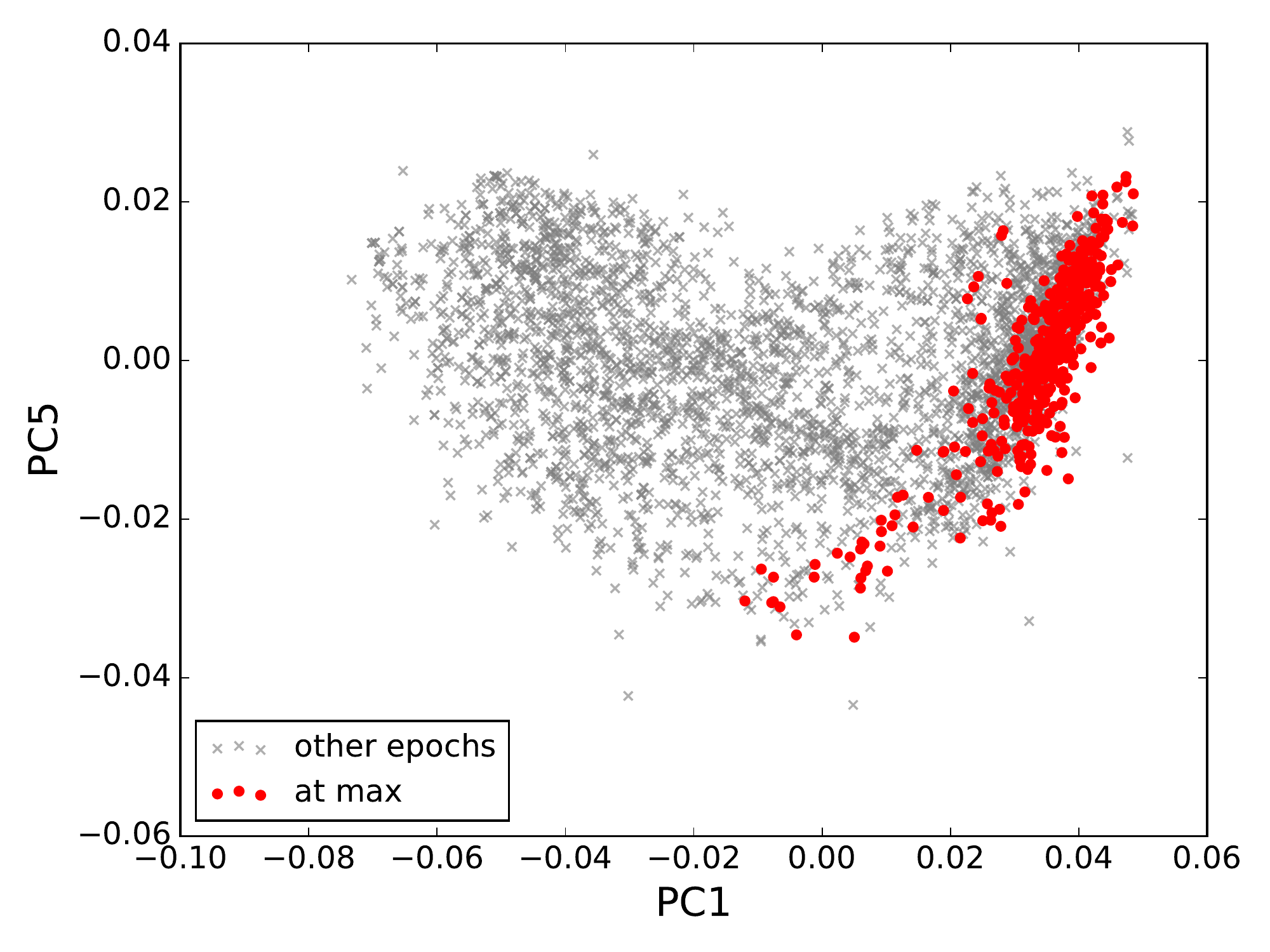}
\caption{Distribution of spectra in the feature space formed by the 1$^{\pmb st}$ and $5^{\pmb th}$ principal components. The red circles represent SN\,Ia spectra at maximum and the grey crosses denote SN\,Ia spectra in other epochs.}
\label{fig:TL_PCA}
\end{figure}

\begin{figure}
\begin{minipage}{1.0\columnwidth}
\includegraphics[scale=0.1, width=1.0\columnwidth]{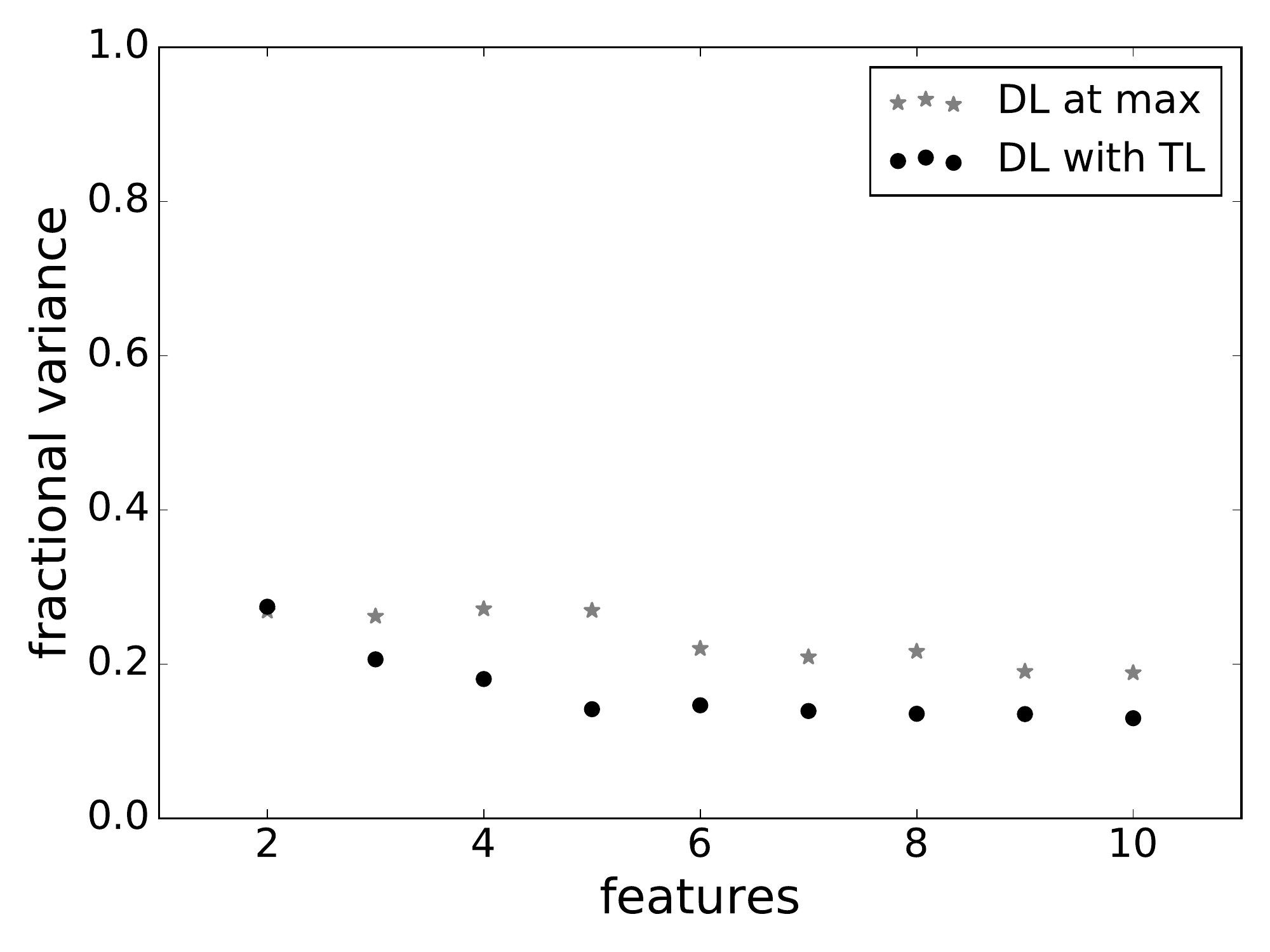}
\caption{Variance between the deep learning reconstructions and observed  SN\,Ia spectra at $B$ max. Gray crosses correspond to a data configuration using only SNe\,Ia at maximum (no transfer learning) and black circles denote results from  an initial data matrix containing spectra from all epochs (with transfer learning). The horizontal axis stands for the number of features; the vertical axis shows the deviation between real data and reconstruction. }
\label{fig:residuals_TL_noTL}
\end{minipage}
\end{figure}

Our data scenario fits within the scope of transfer learning. One is given spectra from the same supernova at various epochs, which can be treated as different observations. Despite being the electromagnetic signature of different astrophysical conditions, spectra from different epochs share common properties (e.g., they all have absorption/emission lines). Consequently, if we consider each spectrum as an independent object (a different line in the data matrix) we can use all of them to train the deep learning network in recognizing spectral features. 

Following this reasoning, our initial data matrix was built with all available SN\,Ia spectra, regardless of the epoch of the observation. This allows us to exploit all available spectra, even those with unknown epoch of  observation, in the investigation of spectra properties at maximum. Closely related strategies, with different goals,  were used by \citet{richards2012} for semi-supervised photometric classification of SN curves, \citet{vilalta2013} for cepheids classification and \citet{kremer2015} for photometric redshift determination. Despite the increase in data volume, the resulting feature space is much more stable, less affected by the inclusion/removal of individual spectra, and provides a safer ground for spectral feature recognition. Thus, each line in our data  matrix holds the derivative of the flux for an observed spectrum between $4000$ and $7000$\AA,  sampled in bins of 10\AA. The complete matrix contains $3677$ lines and 300 columns and serves as input to the dimensionality reduction algorithm (Section \ref{sec:DR}). 

Once the low dimensional space is determined, we select only those spectra of interest (within 3 days from maximum brightness, resulting in 486 individual spectra) for the unsupervised learning phase. Fig.  \ref{fig:TL_PCA} illustrates how the spectra around maximum occupy a well defined region in the principal component parameter space. This configuration can easily be used to study the time evolution of spectral features, as well as to estimate the epoch of a given spectrum. In this work, we focus 
on recognizing characteristics of SN\,Ia spectra at maximum. 

In Fig. \ref{fig:residuals_TL_noTL}, we show how this approach impacts the reconstruction power of the DL feature space. The vertical axis represents residual variance between the reconstruction and the original data with (black circles) and without (grey stars) the transfer learning approach.
To calculate variance, both data sets (one with all the spectra and another containing only spectra at B-maximum) were randomly split in a training  ($80\%$) and a test set ($20\%$). DL  was  applied to both training sets with different number of features in the central layer. Resulting features were than used to reconstruct the spectra in the test set and the normalized residual variances between the predictions and the measured derivative spectra in the test set were then calculated. 
We observe that with transfer learning, 4 features suffice to converge to a stable fractional variance. Without transfer learning, the same performance level cannot be achieved, even if we employ 10 features along the intermediate layer.

\afterpage{
\begin{figure*}
\begin{minipage}{1.0\textwidth}
\centering
\includegraphics[width=0.9\textwidth]{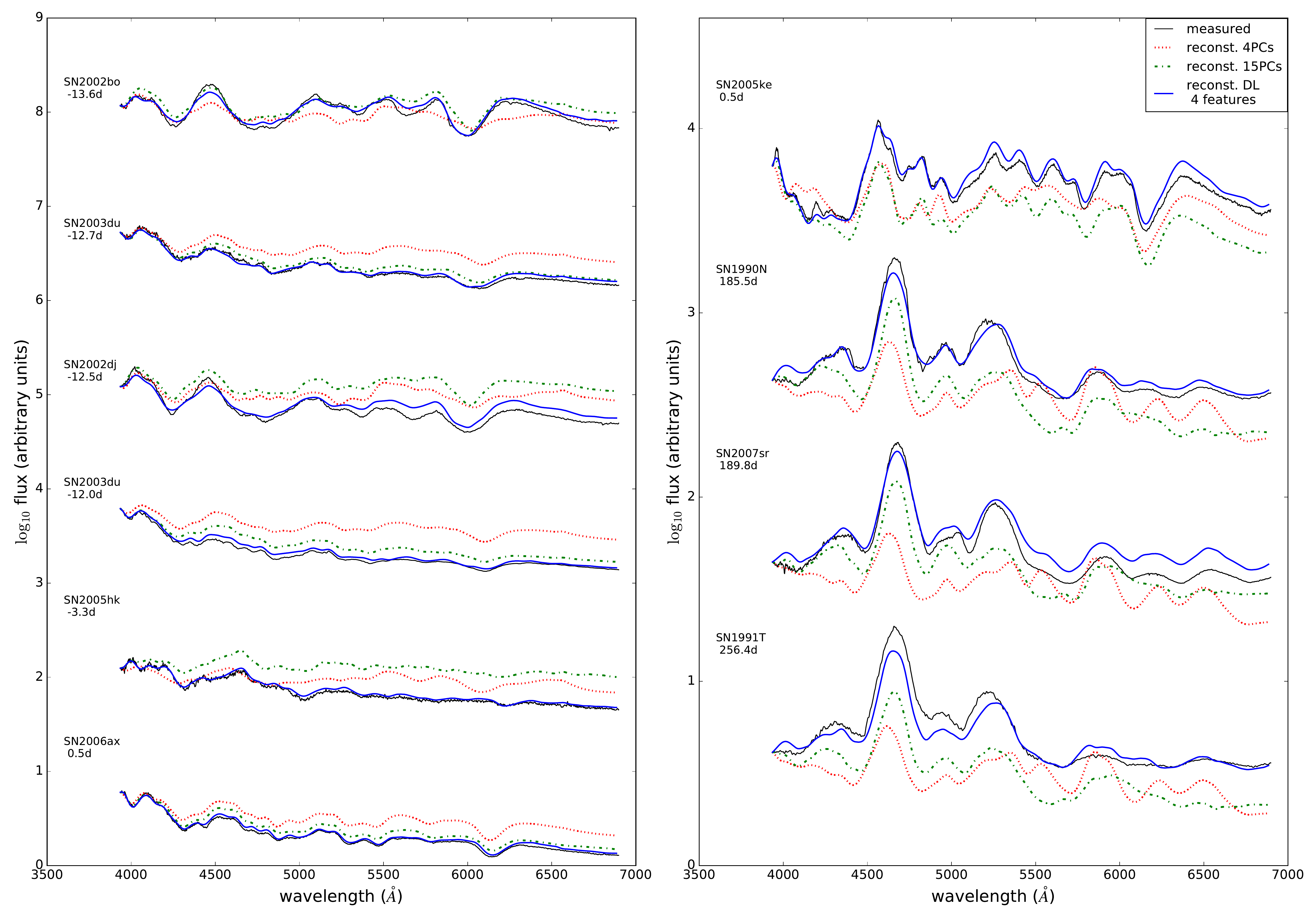}
\caption{Reconstruction of a few examples of measured SN\,Ia spectra (black) using Deep Learning (thin blue) and PCA using 4 (dot-dashed red) and 15 (dashed green) PCs. 
}
\label{fig:rec_examples}
\end{minipage}
\end{figure*}
}

\section{Comparing feature spaces}
\label{sec:comp_ps}

In order to demonstrate how DL outperforms the standard PCA algorithm, we provide  a detailed comparison between the two parameter spaces. Fig. \ref{fig:rec_examples} shows the reconstruction of the original spectra using PCA and DL. In black we show a few examples of SN\,Ia spectra at three representative epochs. The first four spectra are at ${\sim-13}$ days from {$B$ max}. Three spectra are close to $B$ max and, finally, three spectra are found in the nebular phase at ${\sim+180}$ days from maximum (from top left to bottom right).

Reconstructions using DL show very good agreement with observations at all epochs due to the non-linearity of its representations. It performs exceptionally well in the earliest and the latest spectra and on SNe with rare spectroscopic peculiarities such as, for example, SN\,2005hk. The behaviour of the High Velocity Feature \citep{2005ApJ...623L..37M} of the Si {\sc ii}\,6355\AA\ in the early spectra (for example SN\,2002dj and SN\,2002bo) are also finely reproduced by DL when compared to PCA.
The latter is competitive only away from the early and late epochs, and on objects that are not too peculiar. For PCA to obtain reconstructions comparable to DL, we would need a large number of components. From this, it is clear that DL has an outstanding performance when compared to PCA, even when the latter uses almost 4 times the number of parameters.

\begin{figure}
\begin{minipage}{1.0\columnwidth}
\includegraphics[scale=0.1, width=1.0\columnwidth]{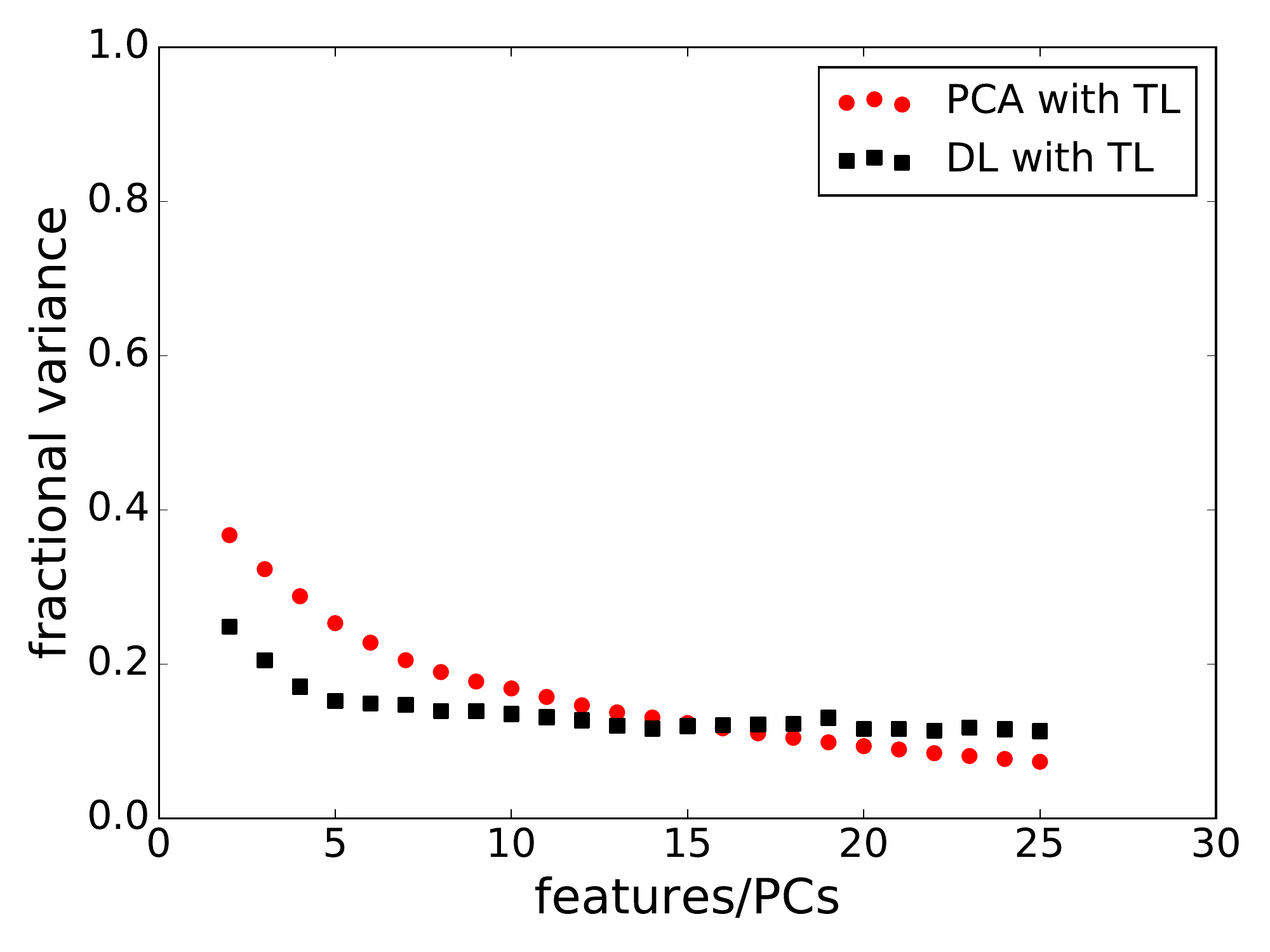}
\caption{Comparison between Deep Learning  and PCA in their capacity to reconstruct the original spectra and robustness to overfitting. Horizontal axis stands for the number of PCs/features and vertical axis shows the deviation between real data and reconstruction. }
\label{fig:residuals_PCA_DL}
\end{minipage}
\end{figure}

Fig. \ref{fig:residuals_PCA_DL}  confirms this superiority quantitatively and, at the same time, highlights still another important advantage of the DL approach: the asymptotic behaviour of residual variances as the number of features increases.
 DL greatly outperforms PCA even when using a small number of features. Its reconstruction capability shows steady improvement until $\sim5$ features, after that it remains approximately constant. 
PCA only achieves a comparable reconstruction with $\sim15$ PCs. However, a large fraction of the variance explained by PCA can be traced to noise, and an unnecessary large number of components overfits the data (this is represented by the constant decrease in fractional variance as a function of the number of PCs). DL behaves robustly, less affected by noise, and preventing overfitting (the variance explained remains approximately constant for more than $\sim$4 features). This suggests that the intrinsic dimension in SN\,Ia spectra is $\sim 5$, leaving only $4$ hidden physical parameters to characterize the explosion (one of these dimensions is needed to explain the time evolution of the spectrum). Similar results concerning the number of significant parameters necessary to describe the spectral features of SN Ia were reported by  \citet{2015MNRAS.447.1247S}. This hints to the apparent simplicity of the space of SN\,Ia spectra and should be considered in model development process.

Once we have demonstrated the superiority of DL in reducing the dimensionality of the parameter space we move to the visualization and unsupervised clustering steps. In what follows, all analyses where performed on the 4 dimensional DL feature space.

\section{Data visualization}
\label{sec:vis}

Given that DL is a relatively new technology in machine learning, and has no precedence
in the study of SN\,Ia spectra, we provide the reader with a couple of visualization tools to enable analysis of this feature space. The two algorithms described below are part of the field of dimensionality reduction, but here we employ them as visualization tools. 

\subsection{Self-organizing maps}
\label{subsec:SOM}
\newcommand{\norm}[1]{\left\lVert#1\right\rVert}

\emph{Self-Organizing Maps} (SOM) are a special kind of \emph{artificial neural networks}, often invoked to visualize data in an unsupervised manner. They are 
commonly used to find a two-dimensional embedding of the data and have
been extensively employed in astronomy, e.g., in stellar spectra \citep{mahdi2011}, light curve classification \citep{brett2004}, and object selection and photometric redshift estimation \citep{way2012, geach2012}.

The main idea behind SOM is to construct a 2-dimensional representation of the data where similar objects are placed close to each other. 
The algorithm can be described as follows: consider data in $\mathbb{R}^d$, and a two-dimensional grid $M = S_1 \times S_2$.  
Initially each cell of the grid, $C \in M$, hosts a random vector $\mathbf{c} \in \mathbb{R}^d$, called prototype (these are initially random,  they iteratively become more representative of the data vectors assigned to their cell). 
One then chooses, at random, one element from the data set, $\mathbf{x} \in \mathbb{R}^d$, and compares it to all prototypes in the grid using, for example, the Euclidean distance in $\mathbb{R}^d$. Let $\mathbf{p}$ be the prototype in the grid that is closest to $\mathbf{x}_i$ and $P \in M$ denote the grid cell hosting $\mathbf{p}$.
Subsequently, $\mathbf{p}$ and all prototypes $\mathbf{q}$ of neighbouring cells $Q$ of $P$ are updated according to the following rule:
\begin{equation}
 \mathbf{q} = \mathbf{q} + \alpha \cdot h(P,Q) \cdot (\mathbf{x} - \mathbf{p}),
\end{equation}
where $h$ is a neighbourhood function (typical functions are $h \equiv 1$, $h(P,Q) = \norm{P-Q}$, or $h(P,Q)=e^{- {(\sqrt{2}\sigma)}^{-2} \norm{P-Q}^2}$) and  $\alpha$, named \emph{learning rate}, is usually reduced during the iterative process. Thus, prototypes of $P$ and its neighbouring cells are made ``more similar'' to $\mathbf{x}$. 
Another element from the data set, $\mathbf{x}$\textquotesingle, is then compared to this new grid configuration. If this data vector is very similar to the first, it is allocated in $P$ or one of its neighbouring cells. Otherwise, it will populate another cell, $P$\textquotesingle, defining a new \textit{locus} on the grid which will host its characteristics. The comparison is repeated for all objects in the data and for the entire data set until convergence. As a result, we are left with a 2-dimensional re-organization of the initial data, where similar objects are allocated in nearby cells, and distinct ones occupy different extremes of the grid. 

\begin{figure*}
  \begin{adjustbox}{addcode={\begin{minipage}{\width}}{%
      \caption{%
       A self organized map of SN\,Ia spectra at maximum, constructed from the Deep Learning 4-dimensional feature space. Black lines correspond to the mean spectra of each cell, purple and blue bands correspond  $1\sigma$  and $2\sigma$ respectively. Also shown are the number of spectra allocated in  each individual cell and the subtype of the majority of SNe populating each cell according to the classification proposed by \citet{Wang09}. In case there are exactly the same number of objects of different subtypes both labels are shown.}
      \label{fig:SOM}
      \makeatletter%
      \ifx\caption@@make\undefined%
        \typeout{\string\caption@@make is undefined, prob. not using \{caption\}}%
      \else%
        \typeout{\string\caption@@make is defined, prob. using \{caption\}}%
        \typeout{\string\caption@box is: \meaning\caption@box}%
      \fi%
      \typeout{\string\@makecaption is: \meaning\@makecaption}%
      \makeatother%
      \end{minipage}},rotate=90,center}
    \includegraphics[width=1.25\textwidth]{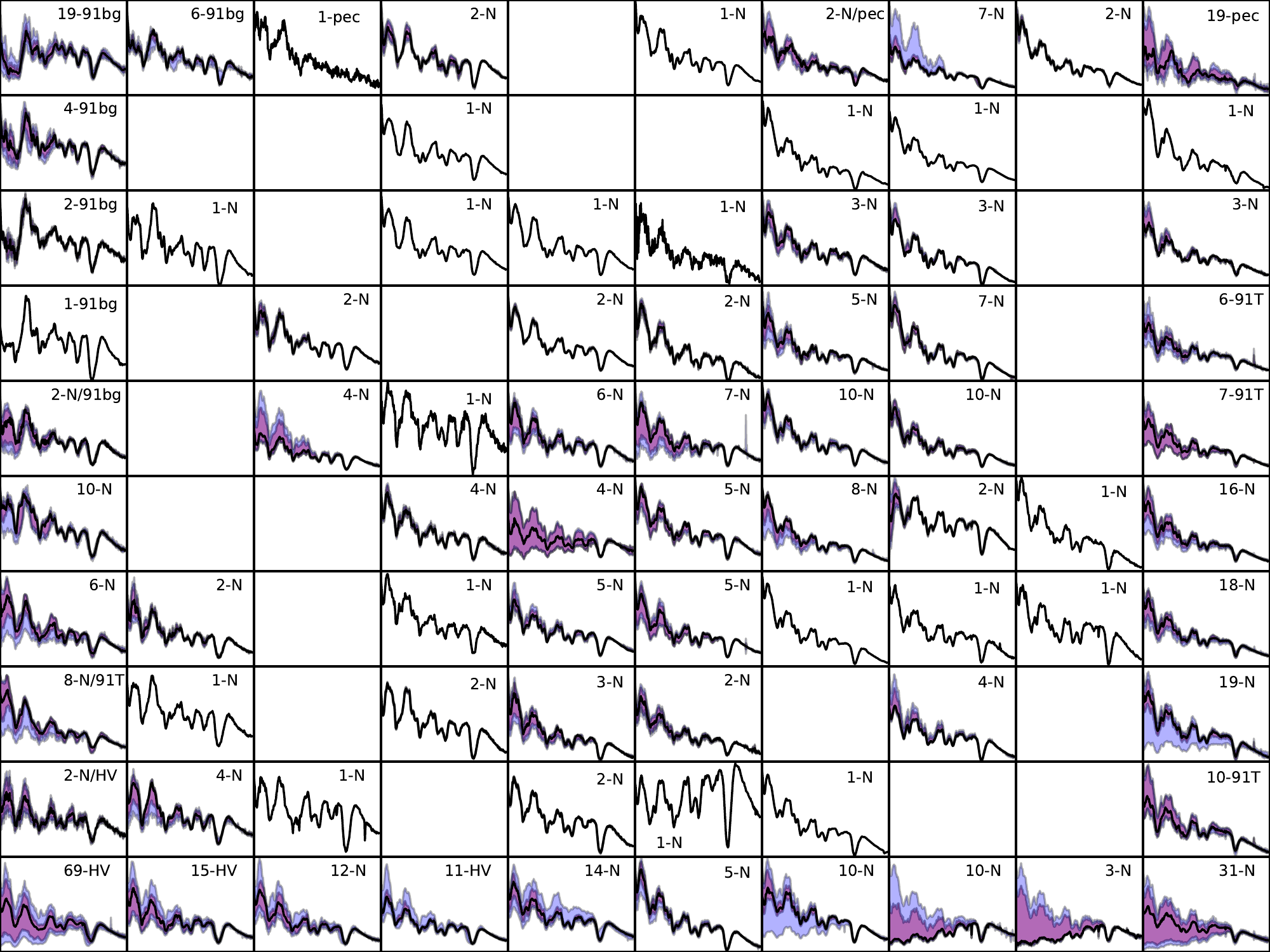}%
  \end{adjustbox}
\end{figure*}

Fig. \ref{fig:SOM} shows results after applying SOM to the 4-dimensional DL  feature space. As described in Section \ref{sec:TL}, in this task we select only spectra close to maximum.
The grid contains 10$\times$10 individual cells showing the mean spectra (black line), standard deviations (pink/purple), the numbers of spectra allocated in each individual cell and the sub-type of the majority of objects in each cell (when the cell population is exactly split between subtypes both labels are shown). We emphasize that Fig.  \ref{fig:SOM} does not show the final prototypes in each cell, but the mean of all the spectra allocated in them. We chose this visualization because the prototypes in our case would relate to the 4-dimensional DL space, making it impossible to recognize spectral features.
Different spectroscopic classes are arranged over different parts of the grid. 
Normal SNe\,Ia such as SN\,1994D and SN\,2011fe are close to the centre of the grid, the peculiar and faint 91bg-like cluster on the top left  and  the 
high velocity SNe are on the bottom left corner.
We also recognize 91T-like SNe on the right side of the grid. From this configuration, we can already tell that the DL feature space is able to grasp crucial differences between different spectral features. 
Literature suggests that 91T-like, High Velocity and 91bg-like SNe are the extremes of SN\,Ia spectral variability, while 91bg-like SNe are possibly a more isolated group \citep{2011MNRAS.410.2137C, 2012AJ....143..126B}.
In this context, spectroscopically normal SNe\,Ia form the bulk of the data set, connecting the other subtypes together through an almost continuous change in spectral features.
These findings are nicely confirmed by our SOM analysis, which gives us a glimpse of the potential of this feature space. In the following sections, we  show how this first visual analysis is in concordance with the more quantitative results obtained using unsupervised learning.

\subsection{Isomap}
\label{subsec:isomap}

Isomap belongs to a broader class of dimensionality reduction techniques known as manifold learning. While PCA seeks to preserve the variance of the data, Isomap preserves its intrinsic geometry~\citep{Tenenbaum2000}. More precisely, it can be seen as an extension of another classical dimensionality reduction method, called \emph{multi-dimensional scaling}~(MDS), which aims at finding a low-dimensional embedding of the data, such that the distance between any pair of two points is preserved. Isomap generalizes this idea by resorting to ``geodesic manifold distances'', which are approximated via, e.g., a neighbourhood graph in which two points (nodes) are connected if one of the points is within the set of the $K$-nearest neighbours of the other one~\citep{Tenenbaum2000}. The geodesic distance, $d_M(i,j)$, between two points $i$ and $j$ can be defined as the shortest path between the two points in that graph. These distances are then used as in classical MDS, which usually resorts to the standard Euclidean distance (``straight lines''). Thus, in contrast to MDS, Isomap can also capture non-linear manifold structures.
In astronomy, it was recently applied to spectroscopic classification by \citet{bu2014}.

In what follows, we use Isomap to provide a 2-dimensional visualization of the 4-dimensional feature space obtained with DL. It provides a much clearer view of the distribution of points in the DL feature space and facilitates a visual comparison with other classifications schemes.

\section{Unsupervised learning}
\label{sec:cluster}

In previous sections we introduced efficient dimensionality reduction techniques. 
We now turn to the last step of our endeavour: unsupervised learning.
Our main goal is to show the feasibility of automatically identifying subtypes  of SNe\,Ia with minimum assumptions about the physics and dynamics of the SN mechanism. Unsupervised learning techniques identify clusters in a data set by maximizing the similarity among objects within the same cluster, and maximizing the dissimilarity among objects from different clusters. 

The computational complexity of a clustering algorithm increases as the dimension of the data grows larger. This is because added dimensions quickly increase the volume of the feature space and the data becomes sparse; the capacity of finding clusters then deteriorates. This effect is known as the \emph{curse of dimensionality}. DL represents our solution to mitigate this effect.

While there are many clustering algorithms readily available, K-Means is certainly the most popular. We have compared different methods using simulated data and found K-Means to exhibit good performance. We focus on this method in the following sections.

\subsection{K-means}
\label{subsubsec:kmeans}

The K-means algorithm is one of the most well-known clustering techniques, yielding intuitive solutions. In its original form \citep{MacQueen1967} it begins by choosing at random $k$ vectors  of the same dimensions of the data (with $k$ chosen by the user). These will act as centres for potential clusters definition. A distance (Euclidean) is then calculated between  all vectors in the data set and the centre candidates. Each data point is assigned to the cluster represented by its closest centre candidate. Once the first set of clusters is defined, the centre candidates are updated to the centroid defined by all the members in each cluster. The process is repeated until the centroids are not changed due to further iterations. 
In practice, this local search strategy quickly converges to a solution (e.g., after 50 iterations). 

One drawback of K-means is the need to explicitly state the number of clusters \emph{a priori}. In our case, we wish to show that our approach is able to grasp the main spectral features underlying the classification proposed by \citet{Wang09} (which is composed by subtypes normal, high-velocity, 91T-like and 91bg-like) and, as a consequence, we will 
search for up to 4 clusters. A deeper analysis quantifying the degree of cluster coherence throughout different number of clusters will be addressed in a subsequent  paper. 

All the methods described above have been made available in a single toolbox that enables quick analysis of a (potentially) large initial data set. We present \texttt{DRACULA} (Dimensionality Reduction And Clustering for Unsupervised Learning in Astronomy) in detail in appendix \ref{sec:code}.

\section{Results}
\label{sec:results}

\begin{figure*}
\begin{minipage}[b]{0.9\textwidth}
\centering
\includegraphics[width=0.9\textwidth]{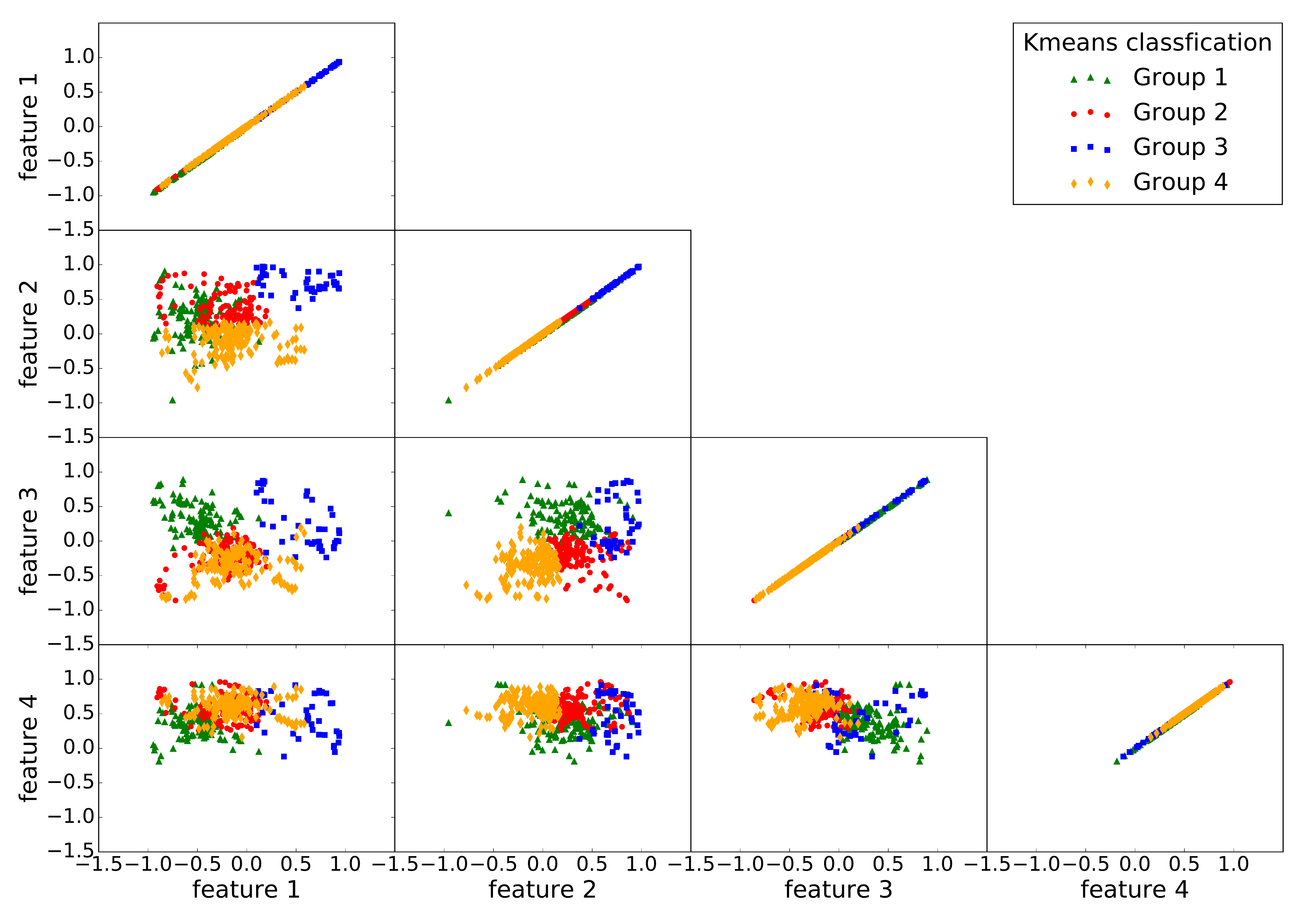}
\end{minipage}
\newline
\begin{minipage}[b]{0.9\textwidth}
\centering
\includegraphics[width=0.9\textwidth]{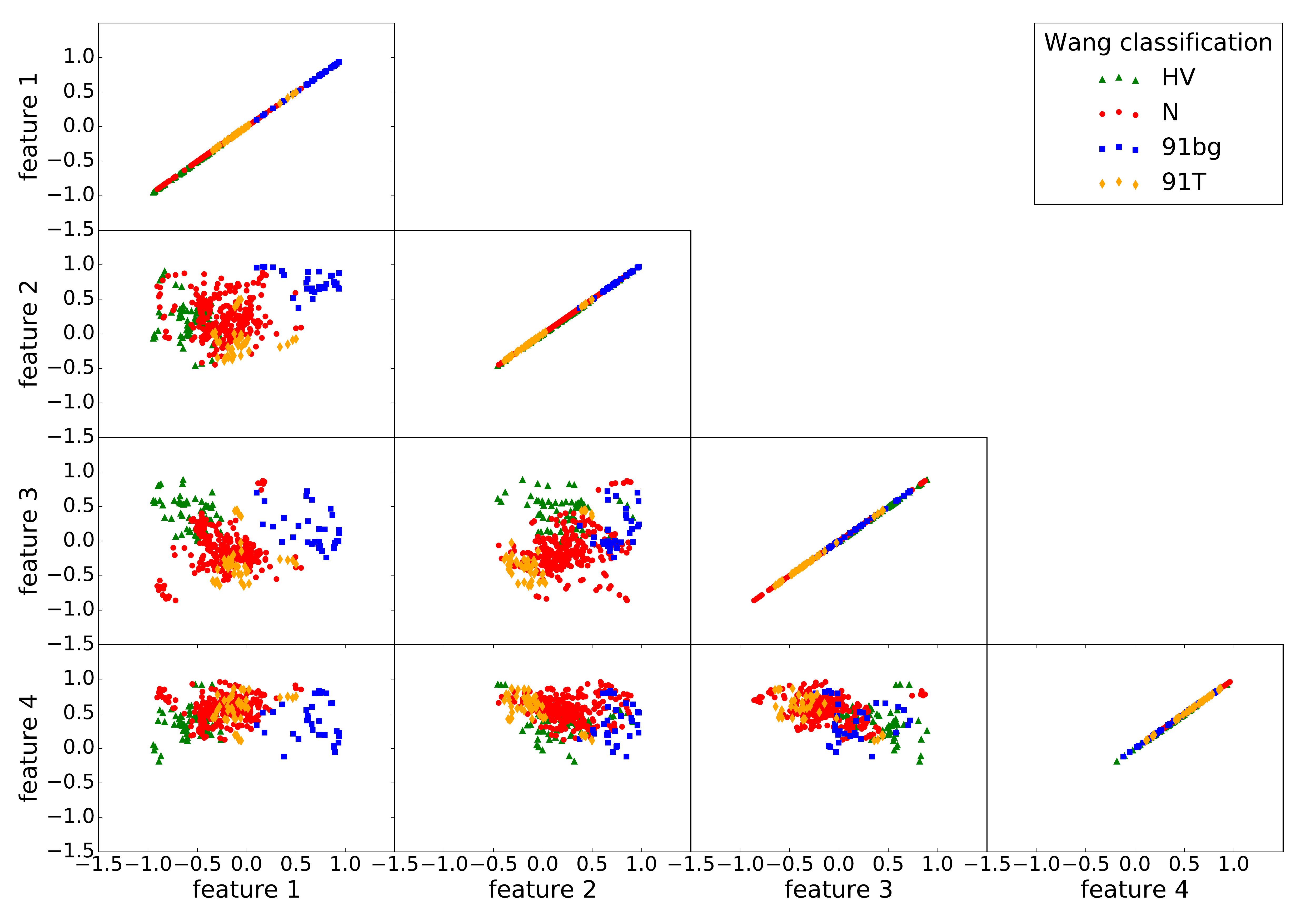}
\caption{Four dimensional feature space resulting from Deep Learning. Each point corresponds to one SN\,Ia spectrum within $-3$ and $+3$ days since $B$-maximum. \textbf{Top:} Colours represent the groups found by the K-Means algorithm when 4 groups are required. \textbf{Bottom:} Colours correspond to the classification suggested by \citet{Wang09}. This plot does not show the objects classified as ``peculiar''.}
\label{fig2:DL_KM_wang_scatter}
\end{minipage}
\end{figure*}

We used \texttt{DRACULA} to analyse the sample of SN\,Ia spectra introduced in Section \ref{sec:data}. Results shown below correspond to the 4-dimensional DL feature space (Section \ref{sec:comp_ps}) ran through the K-Means algorithm. In order to compare our results with the classification proposed by \citet{Wang09}, we set K-Means to search for 4 distinct clusters (Section \ref{sec:cluster}). 

Fig. \ref{fig2:DL_KM_wang_scatter} shows the 4-dimensional DL feature space configuration. In the upper panel, colours correspond to the clusters found by K-Means; in the lower panel points are identified according to \cite{Wang09} classification (the lower panel of Fig. \ref{fig2:DL_KM_wang_scatter} does not show the SNe classified as ``peculiar'' by \citet{Wang09} because there is not a unique underlying spectral  characteristic which defines this group). Although the two classification schemes are not in complete agreement, they share some basic characteristics. For example, the scatter plot in the feature space formed by features 1 and 2  are quite similar in the upper and lower panels, indicating the correspondence between group 3 (upper panel) and 1991bg-like SNe (lower panel). 

\begin{figure*}
\begin{minipage}{0.9\textwidth}
\centering
\includegraphics[width=0.9\textwidth]{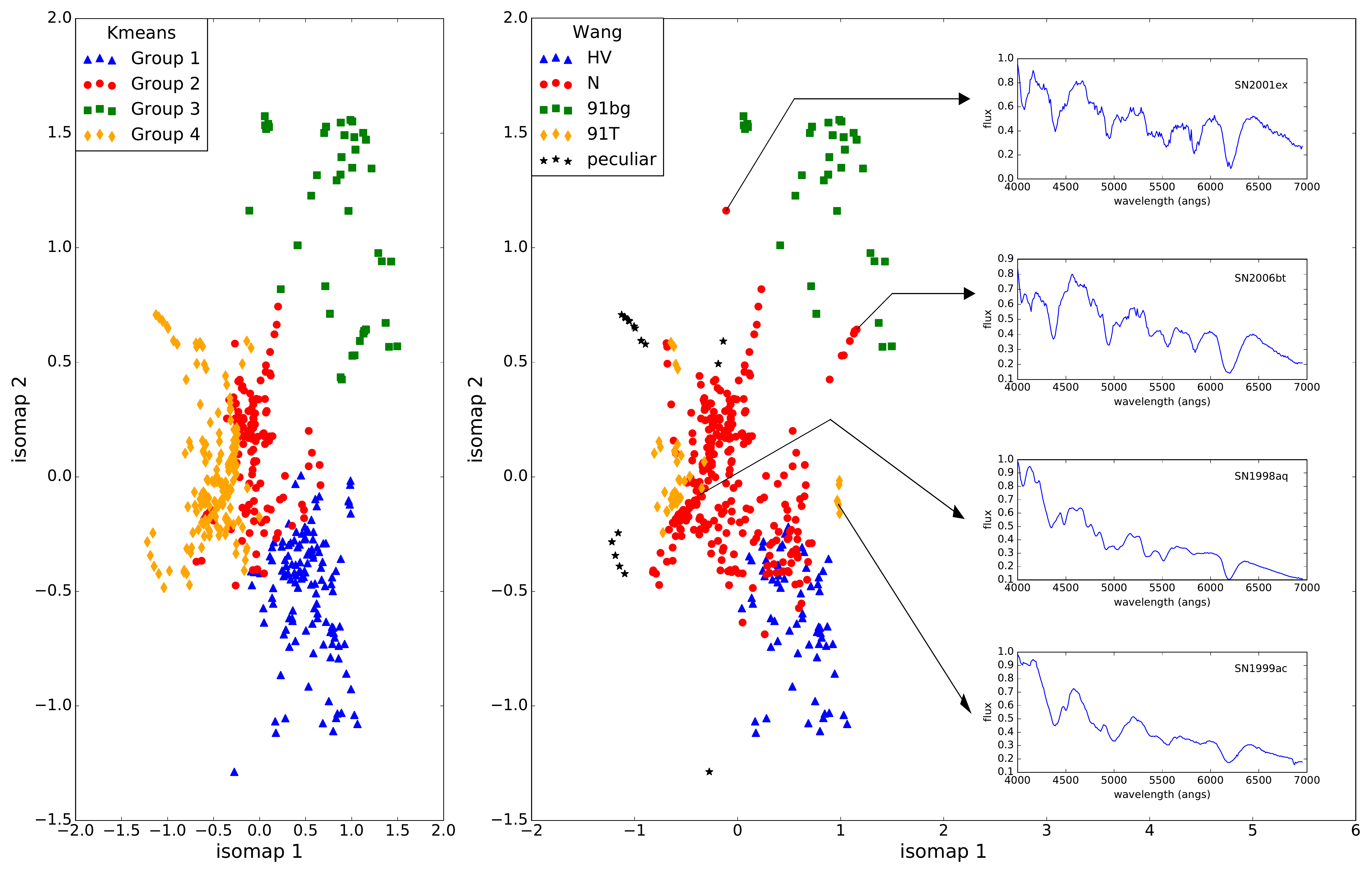}
\caption{Four dimensional feature space from Deep Learning reduced to 2 dimensions through isomap. \textbf{Left:} Groups found by the K-Means algorithm when 4 groups are imposed. \textbf{Right:} Objects separated according to the classification proposed by  \citet{Wang09}. In this panel we also highlight spectra of 4 SNe lying in the boundary between classes.}
\label{fig:2d_deep}
\end{minipage}
\vfill
\begin{minipage}{0.9\textwidth}
\centering
\includegraphics[width=0.9\textwidth]{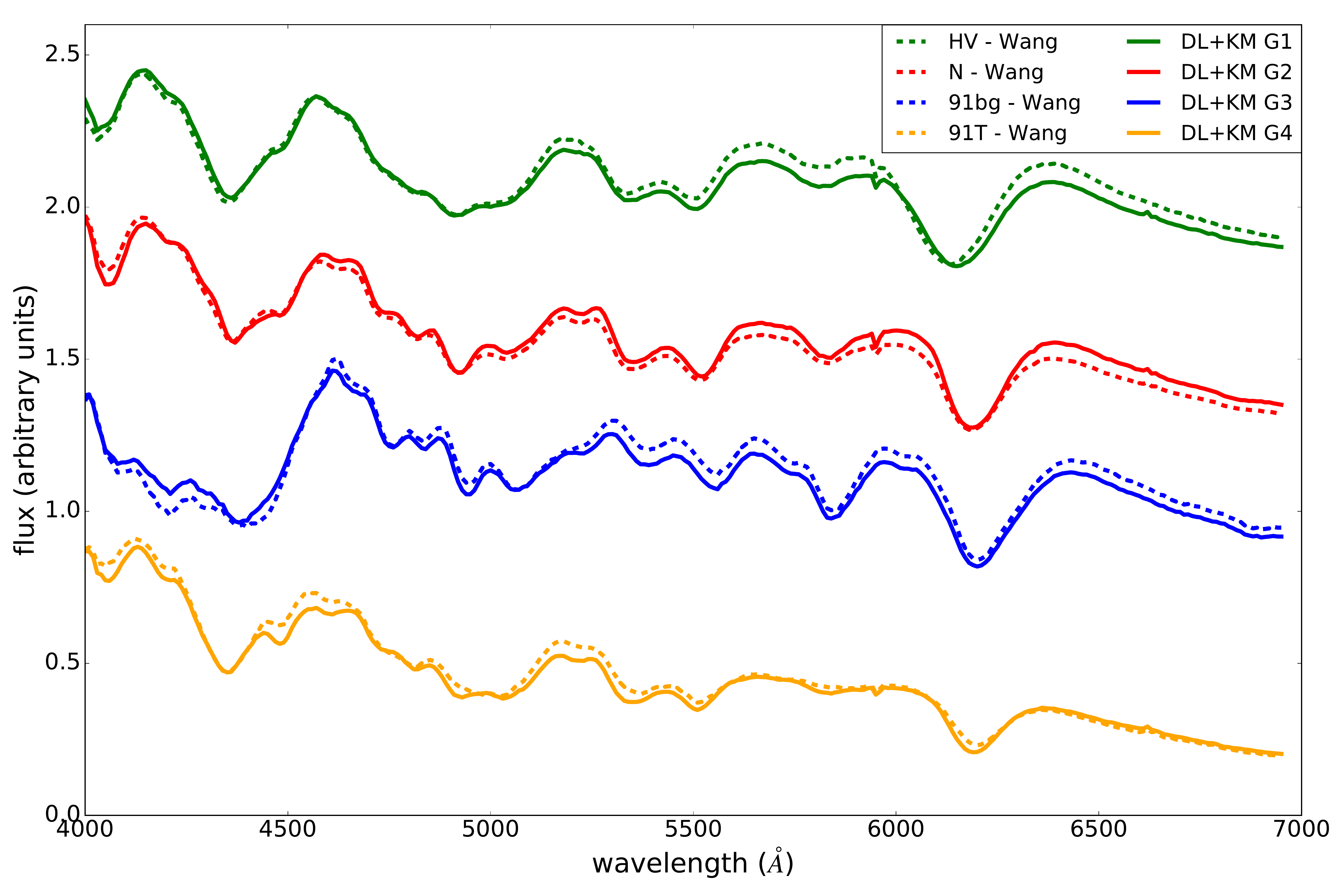}
\caption{Comparison between the mean spectrum found by Deep Learning + K-Means (full lines) and mean spectrum of the SNe Ia subtypes defined by  \citet[][dashed lines]{Wang09}.}
\label{fig2:wang_DL_km_spectra}
\end{minipage}
\end{figure*}

A better visualization of this feature space is achieved by applying the isomap algorithm (Section \ref{subsec:isomap}) to the 4-dimensional DL feature space. Fig. \ref{fig:2d_deep} shows the resulting 2-dimensional isomap space with the clusters found by K-Means (left panel) and the SN\,Ia subtypes defined by \citet[][right panel]{Wang09}. The spectra of a few SNe located at the transition between two subclasses are highlighted. This figure not only demonstrates how isomaps can be a powerful tool in high dimensional data visualization, but also clarifies the potential of combining DL with unsupervised learning  algorithms. Moreover, it confirms  that currently defined SNe Ia subtypes are extremes of a continuous distribution of spectral features.
Many of the SNe located between different classes have been recognized as peculiar  
\citep[e.g. SN\,2006bt][]{2010ApJ...708.1748F}, or ``unusual'' \citep[SN\,1999ac][]{2005AJ....130.2278G} objects,
and the classification of SN\,1999ac as a 91T-like is at least dubious \citep{2006AJ....131.2615P}. Peculiar characteristics, like a blue $B-V$ colour typical of 91T-like objects, are recognized even in the cases where these transitional objects were classified as normal
 \citep[e.g. SN\,1998aq][]{2003AJ....126.1489B}. This is a clear evidence of the existence of intermediary objects in the frontiers of SNe\,Ia subclasses and the consequent continuum characteristics of its spectral features space.

Our goal is to identify which spectral characteristics  correspond to the extremes. 
Fig. \ref{fig2:wang_DL_km_spectra} shows the mean spectrum of each cluster in both classification schemes. The agreement between the mean spectra is proof that DL, when coupled with unsupervised learning algorithms, is able to automatically identify important spectral features without human screening. The method recovers classes similar to those defined by visual inspection; this can be used to optimize the current identification of types or subtypes of SNe. 

Our methodology is also capable of identifying a hierarchical structure within the data set. Fig. \ref{fig:KMeans1234} shows the mean spectrum of all SN\,Ia spectra at maximum (top-left panel) as well as the mean spectrum of each cluster found by K-Means using $k=2$ to $k=4$ clusters, comparing it with the mean spectra of subtypes proposed by \citet{Wang09}. Given the data set at hand, we see that the velocity at which most lines form (the velocity of the photosphere) is the first order spectral characteristic which defines subtypes of SNe\,Ia (2-cluster configuration). After this, 1991bg-like objects are kept in a cluster of their own (3-cluster configuration) and finally 91T-like objects are separated. Depending on the degree of specialization we demand from the data, we are able to recognize a sequence of spectral features which might be used to guide the physical basis of future data-driven classification systems.  

Our analysis suggests that the currently defined SN\,Ia subtypes are extremes of a continuous distribution of spectral features and that SNe\,Ia  live in a small multi-dimensional continuum with no strict boundaries separating subclasses and likely governed by few key physical parameters.

\afterpage{
\begin{figure*}
\centering
\begin{minipage}{1.0\textwidth}
\centering
\includegraphics[width=1.0\textwidth]{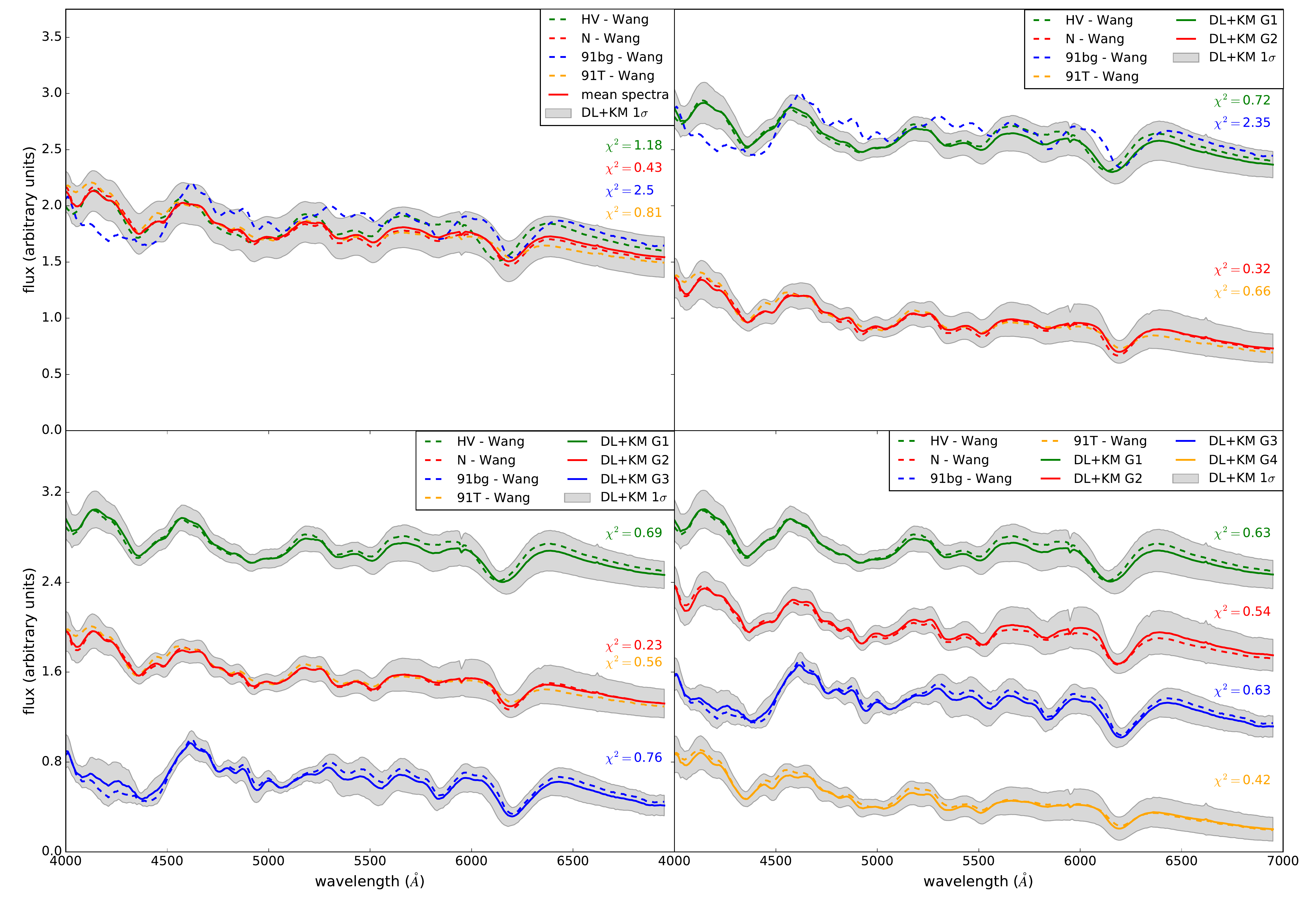}
\caption{Representation of the hierarchical evolution of the groups found by K-Means algorithm (full lines) in comparison with the groups defined by \citet{Wang09} (dashed lines). The top-left panel shows the mean of all SN\,Ia spectra between $-3$ and $+3$ days since B-maximum (full red line). Other panels show the mean spectra for different configurations of the K-Means algorithm, from 2 to 4 groups. Also shown are the deviations between the mean spectra between groups identified with the K-Means algorithm and the mean spectra for the groups defined by \citet{Wang09}. The grey region denotes $1\sigma$ scatter for the groups defined with the K-Means algorithm.}
\label{fig:KMeans1234}
\end{minipage}
\end{figure*}
}

\section{Summary and Discussion}
\label{sec:discussion}

We propose a framework to automatically identify subtypes of SNe Ia within a set of measured spectra by combining modern machine learning techniques. As a first application of this tool, we investigate how to recover previously reported subtypes of SNe\,Ia. 

The set of public SN\,Ia spectra (Section \ref{sec:data}) was first submitted to the preprocessing described in \cite{2015MNRAS.447.1247S} and the resulting matrix was used as input for the algorithm which can be summarized in 3 main steps: transfer learning, dimensionality reduction, and unsupervised learning. 

The goal of transfer learning is to ensure the stability of the low dimensional feature space by adding a large variety of spectra in the original data matrix. 
In our example, although the clustering analysis is focused at B-maximum, we use spectra from all available epochs for training. Once the low dimensional feature space is constructed, only the projections corresponding to spectra at maximum are selected for the next phase (figure \ref{fig:TL_PCA}).

We introduce Deep Learning for dimensionality reduction on SN\,Ia spectra. This is a cutting-edge technique only recently introduced to astronomy \citep{company2015}. We prove its effectiveness in spectroscopy data analysis and show that it outperforms the PCA algorithm in the reconstruction of measured spectra, reducing the dimensionality of the feature space from $\sim 300$ hundreds to 4. Since this is the first application of Deep Learning for SN\,Ia spectra characterization, we use \textit{Self- Organizing Maps} (SOM) to better understand the potential of the new reduced feature space (figure \ref{fig:SOM}); this visualization technique shows how spectral properties vary in the SN\,Ia spectra space; specifically it allows the visualization of the peculiarity of subtypes reported in the literature, e.g., the 91bg-like SNe \citep{2011MNRAS.410.2137C,2012AJ....143..126B}.

Lastly, we use unsupervised learning techniques to investigate the possibility of identifying spectroscopic features in subclasses of SN\,Ia spectra at maximum light. 
This allows us to define a data-driven classification scheme \textit{a posteriori} and analyse clusters separately in order to look for their spectroscopic characteristics. This facilitates the classification of a fairly large data set requiring the astronomer to visually inspect only a handful of possibilities (the mean spectra). 
We use the low dimensional space from Deep Learning as an input for the K-Means algorithm. In order to provide a more friendly visualization of the four dimensional feature space, we use isomap as a further layer of dimensionality reduction. Here the separation between the clusters is  more evident; the identification with the \citet{Wang09} clusters is also clearer. 

We find that the spectral variability of SNe\,Ia can be summarized by a low dimensional space. Spectra starting from few days from the explosion up to more than a year afterwards can be parametrized by a 5-dimensional space. The time evolution of the spectra uses one of these dimensions. This result suggests that only 4 underlying physical parameters suffice to describe SN\,Ia explosions and their spectroscopic variability, including the most ``peculiar'' objects, such as 91bg-like and the 02cx-like SNe. 

SNe\,Ia are well known as a uniform class of objects. Our results prove this claim, and suggest that progenitors should be a ``simple'' system with no more than a handful of initial parameters. Moreover, we also show that the currently identified SN Ia sub-types are in fact the extremes of a continuous distribution of spectral features. We also do not find strong evidences of distinct subclasses.

Our complete software apparatus was built under \texttt{DRACULA}, a publicly available Python package that is here tested on a public data set of SN\,Ia spectra. Our results show agreement between mean cluster spectra and those proposed by \citet{Wang09}. Our method is also able to identify a hierarchical structure within the data set, confirming previous statements that high velocity features are of the first order effect in separating currently available samples of SNe\,Ia \citep{wang2013}. Future work considers analysis of time evolution in SN\,Ia spectra, and the possibility of using our tool in the classification of other supernova types.

\section*{Acknowledgements}

We thank Bruce Bassett, Or Graur, Paniez Paykari and Zoe Vallis for insightful discussions and comments. EEOI and RSS thank  Petr Skoda for pointing out the capabilities of SOM. EEOI is partially supported by the Brazilian agency CAPES (grant number 9229-13-2).
MA is supported by S\~{a}o Paulo Research Foundation (FAPESP) under grant number 2013/26612-2.
VCB is supported by S\~{a}o Paulo Research Foundation (FAPESP)/CAPES agreement under grant number 2014/21098-1. HC is supported by CNPq under grant number 141935/2014-6. YF is supported by the ERC grant Pascal 277742. AMMT was supported by the FCT/IDPASC grant contract SFRH/BD/51647/2011. This work is a product of the 2$^{\rm nd}$ COIN Residence Program. We thank Alan Heavens and Jason McEwen for encouraging the realization of this edition.  The program was held in the Isle of Wight, UK in October/2015 and supported by the Imperial Centre for Inference and Cosmology (ICIC), Imperial College of London, UK, and by the Mullard Space Science Laboratory (MSSL) at the University College of London, UK. 
The IAA Cosmostatistics Initiative\footnote{\url{https://asaip.psu.edu/organizations/iaa/iaa-working-group-of-cosmostatistics}} (COIN) is a non-profit organization whose aim is to nourish the synergy between astrophysics, cosmology, statistics and machine learning communities. This work was written on the collaborative \texttt{Overleaf} platform\footnote{\url{www.overleaf.com}}, and made use of the GitHub\footnote{\url{www.github.com}},  a web-based hosting service, the \texttt{git} version control software, and Slack\footnote{\url{https://slack.com}}, a team collaboration platform.
\appendix
\section{Glossary}
\label{ap:glossary}

To avoid confusion due to different nomenclatures used by the machine learning and astronomy communities, we provide a list of the terms used in this paper:

\begin{itemize}
\item Class/subclass (Supernova Physics): these denote different underlying physical models or characteristics of the progenitor system.
\item Feature (Machine Learning): a recorded property or observation. In our context it corresponds to the flux (or the derivative of the flux) of a given wavelength bin. 
\item Feature space (Machine Learning): commonly known in astronomy as parameter space. This space is originally of high dimensionality ($\sim 300$ wavelength bins), but is reduced to a 4-dimensional space when we invoke Deep Learning.
\item Parameter (Computing): an input variable of a computer algorithm.
\item Physical parameter (Supernova Physics): a generic term to describe one of the initial conditions of the supernova explosion.
\item Prototype (SOM): vectors populating each cell of a SOM grid. Initially random,  they iteratively become more representative of the data vectors assigned to their cell. 
\item Spectral feature (Spectroscopy): absorption and/or emission feature in the spectrum. It originates from a group of atomic lines with similar energy. It is usually dominated by the lines of a single ion.
\item Type/subtype (Supernova Physics): these denote different categories of SNe based on spectral features (e.g. HV, 91T-like, etc.).
\item Weight parameter (Deep Learning): weight matrices connecting adjacent layers of the Deep Learning network. Weights are optimized during the training phase using the training sample.
\end{itemize}
\afterpage{
\begin{figure}
\begin{minipage}{1.0\columnwidth}
\includegraphics[width=1.0\columnwidth, trim={150 265 170 130}]{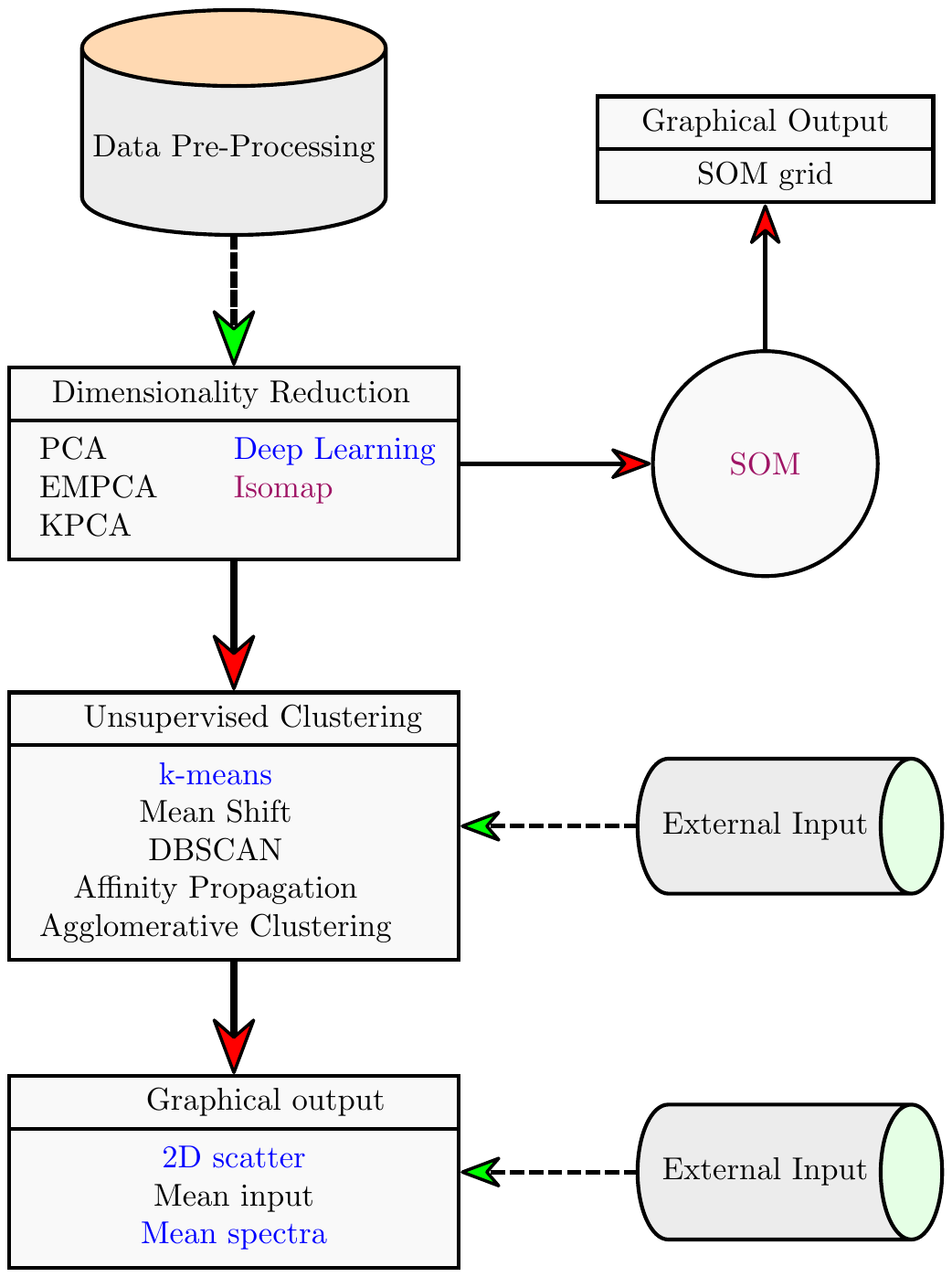}
\caption{Flow chart describing the capabilities of the \texttt{DRACULA} package. Cylinders represent external inputs, rectangles denote package tools, and circles represent independent modules. Red (full) arrows indicate the complete algorithm, and green (dashed) arrows indicate points where external inputs can be inserted. Names marked in blue are the ones used to obtain the results in Section \ref{sec:results}; the ones marked in purple are dimensionality reduction tools used here for visualization only.}
\label{fig:flowchart}
\end{minipage}
\end{figure}
}

\section{DRACULA}
\label{sec:code}

We present \texttt{DRACULA} (Dimensionality Reduction And Clustering for Unsupervised Learning in Astronomy), an implementation of the methods discussed in this paper. The toolbox is written in Python, is publicly available, and can be easily adapted to multiple applications. The software relies heavily on tools developed in scikit-learn \citep{scikit-learn}. The DL analysis used the H2O package \citep{h2o} 
and the SOM routine used the \citet{kajic2014} implementation. Thanks to its modular design, all main steps, e.g., dimensionality reduction, unsupervised learning (clustering) and plotting, can be run separately. A flowchart illustrating the code capabilities is shown in Fig. \ref{fig:flowchart}.

In what follows, we demonstrate how \texttt{DRACULA} can be invoked to obtain similar results to those presented in this paper\footnote{If  you want to be updated on \texttt{DRACULA} development send a request to  \url{coin_dracula+subscribe@googlegroups.com}}. 
We avoid a detailed description of all code functionalities, and focus on main procedures (the documentation\footnote{\url{https://github.com/COINtoolbox/DRACULA}}  provides more detailed descriptions). Let us start assuming an initially big data matrix. In our case, this is composed of derivatives over the flux logarithm, as described in Section~\ref{sec:data}. 
We now go through each step described in previous sections and demonstrate how the user can reproduce our results using \texttt{DRACULA}.

\subsection{Dimensionality Reduction}

To begin, the user needs to set up a configuration file (which must be named \texttt{config.py}) to define the type of desired analysis. This module contains four methods: PCA \citep{jolliffe1986}, Expectation Maximization PCA \citep{bailey2012}, Kernel PCA \citep{scholkopf1999} and DL \citep{deng2014}. Dimensionality reduction requires as input a list of observed features for each object (1 line per object, 1 column per feature). For clarity, input and output are defined next: 

\iout
{data amenable to reduction (ex: spectra derivatives)}
{reduced data}

In order to perform dimensionality reduction with DL\footnote{Note that the codes uses \texttt{R} as an interface, which requires installation of \texttt{R} (\url{https://www.r-project.org/}), h2o (\url{http://h2o.ai/}) and rpy2 (\url{http://rpy.sourceforge.net/}) packages.} the configuration file must 
contain the keywords: \texttt{ORG\_DATA}, that receives the path to the file with uncompressed data and \texttt{REDUCTION\_METHOD} that determines the dimensionality reduction algorithm. If no other options are included in the configuration file, the code will run using default values\footnote{Initial random seed: \texttt{DeepLearning\_seed $=$ 1}; number of hidden layers \path{DeepLearning_n_layers = 7}; structure of hidden layers: \path{DeepLearning_hidden =  'c(120,100,90,50,30,20,4,20,30,50,90,100,120)'}}. The user might change default values by modifying the random seed, \texttt{DeepLearning\_seed}, number and structure of hidden layers, \path{DeepLearning_n_layers} and  \texttt{DeepLearning\_hidden} respectively, in the configuration file.

One can run the chosen dimensionality reduction routine from the command line typing DRAC\_REDUCTION.
The output containing the reduced data is created in a folder called \texttt{red\_data}. 

\subsection{Clustering}

Clustering follows using the output file from the last section or using an externally reduced dataset. In the case of an external source, \texttt{ORG\_DATA} should be commented out and the path to the externally reduced file must be given as

\begin{lstlisting}
REDUCED_DATA_EXTERNAL = "<path to reduced data>"
\end{lstlisting}

The running format is the same as dimensionality reduction, where the input is the reduced data. After clusters are detected, the algorithm outputs their centres and the corresponding labels for each object. In short: 

\iout
{reduced data}
{centers of clusters and labels for each object in the reduced data file}

Available options here are KMeans \citep{MacQueen1967, arthur2007}, Mean Shift \citep{comaniciu2002}, Agglomerative Clustering \citep{voorhees1986}, Affinity Propagation \citep{comaniciu2002} and DBSCAN \citep{ester1996}.
For our particular case, the configuration file contains:

\begin{lstlisting}
CLUSTERING_METHOD = "KMeans"
\end{lstlisting}

Analogous to the dimensionality reduction case, the user can change parameters. 
An example is the number of clusters, using variable \path{KMeans_n_clusters}. The default value is \path{KMeans_n_clusters =  4}.
At this stage it is also possible to use a mask, which is specially important for cases where transfer learning is applied (e.g., when the dimensionality reduction algorithm is applied to a big diverse matrix and only a subset of the reduced data is intended for clustering). The mask consists of a file with the same number of lines as the reduced data; in each line ``1'' indicates objects included in the clustering analysis and ``0'' refers to all remaining objects. The path for the mask should be provided as follows: 

\begin{lstlisting}
MASK_DATA = '<path to mask file>'
\end{lstlisting}

The clustering routines can be run in the command line by typing DRAC\_CLUSTERING.

\subsection{Varying parameters}

In cases  where it is necessary to compare the output from a range of parameters, \texttt{DRACULA} allows the user do it automatically, however, beware that this functionality only permits to change one parameter at a time. The parameters to be declared in the configuration file are

\begin{lstlisting}
VAR_TYPE = 'CLUSTERING' or 'REDUCTION'
VAR_PAR = 'REDUCTION_METHOD' or 'CLUSTERING_METHOD' 
VAR_VALS = [1,2,3] or ['name1','name2','name2'] or [vec1,vec2,vec3]
\end{lstlisting}
This functionality uses the same configuration file (\texttt{config.py}) with the above extra keys, and can be run by typing DRAC\_COMPARISON.

\subsection{Plotting}

It is useful to plot results to gain insight on the cluster quality. Plots can be generated for the whole process, from reduction to clustering, with input and output given by:

\iout
{reduced data, cluster centres and labels for each object in the reduced data file}
{scatter plot of reduced data coloured according to labels}

Scatter plots, similar to those shown in Fig. \ref{fig2:DL_KM_wang_scatter}, are generated by typing DRAC\_PLOT. The format of the output files is selected by setting the keyword \texttt{PLOT\_EXT}, in the configuration file.

If one wishes to use this tool for plotting clusters assigned by an external clustering algorithm, the path to the corresponding labels should be provided:

\begin{lstlisting}
LABELS_DATA_EXTERNAL = "<path to labels file>"
\end{lstlisting}

Finally, it is useful to visualize the mean input data  
within each cluster so features can be compared with expected patterns. The interface in this case is

\iout
{reduced data and cluster labels}
{plot of mean input data for each cluster}

If the data for visualization is not in the file used to make the reduction (in our case the reduction is performed on the derivative of the spectra, but we would like to observe the mean spectrum of each group found by K-Means), it can be plotted by setting:
\begin{lstlisting}
SPECTRAL_DATA_EXTERNAL = "<path to original data file>"
\end{lstlisting}

\noindent The format of the output file for the mean data (or mean spectrum) needs to be set separately through the keyword \texttt{PLOT\_SPEC\_EXT}. The plots can be generated using the command DRAC\_PLOT\_SPECS.

We emphasize that the above provides only a glimpse of the capabilities of DRACULA. The package has other functionalities (e.g. cluster validation routines) which will be fully explored, and described, in  subsequent work.

\subsection{SOM}

The SOM module (Section \ref{subsec:SOM}) is independent from the steps previously described. 
The interface is as follows: 

\iout
{reduced data}
{SOM grid}

This module requires its own configuration file, which must be called \texttt{config\_som.py}. The path to the reduced data file is given through the keyword \texttt{ORG\_DATA}. When no other option is provided in the configuration file, the module will run a $10\times10$ matrix through 100 iterations, and the output plot will show the prototypes populating each cell. If the user wants the SOM grid to show the mean of the original data (in our case, the mean of observed spectra assigned to each cell), the keyword \texttt{SPECTRAL\_DATA\_EXTERNAL} must point to the original data file. The number of iterations can be set by adding the keyword \texttt{Niter} to the configuration file. The grid shown in Fig. \ref{fig:SOM} was constructed with \texttt{Niter = 10000000}. SOM module can be run by typing DRAC\_SOM.


\bibliographystyle{mnras}
\bibliography{ref}

\bsp	
\label{lastpage}
\end{document}